\documentclass[11pt,a4paper]{article}
\usepackage[utf8]{inputenc}
\usepackage{textcomp}

\usepackage{graphicx}
\usepackage{amsmath}
\usepackage[version=4]{mhchem}
\usepackage{siunitx}
\usepackage{longtable,tabularx}
\usepackage{authblk}
\begin{document}

\title{Simultaneous Measurement of Circular Dichroism and Circular Differential Scattering}

\author[1]{Qiang Hao\thanks{wlsichuang@163.com}}
\author[1]{Pathum Wathudura}
\author[1]{Huy Pham}
\author[2]{In Han Ha}
\author[3]{Abrahan Martinez}
\author[1]{Justin Lovett}
\author[1]{Nicholas C. Fitzkee}
\author[2]{Ki Tae Nam}
\author[3]{Shengli Zou}
\author[1]{Dongmao Zhang\thanks{dongmao@chemistry.msstate.edu}}

\affil[1]{Department of Chemistry, Mississippi State University, Mississippi State, Mississippi 39759, United States.}
\affil[2]{Department of Materials Science and Engineering, Seoul National University, Seoul 08826, Republic of Korea.}

\affil[3]{Department of Chemistry, University of Central Florida, Orlando, Florida 32816, United States.}

\maketitle

\begin{abstract}
Chiroptical spectroscopy provides a non-invasive, label-free approach for resolving microscopic structural details via interactions with circularly polarized light. Despite the widespread application and complementary information provided for chiroptical materials characterization, the simultaneous acquisition of circular dichroism (CD) and circular differential scattering (CDS) spectra has remained challenging. In this work, we develop a dual-channel spectrometer that enables the acquisition of CD and CDS spectra from the same solution. To address the challenge of CDS baseline correction, we introduce a scattering spectral matching method. The performance of the instrument is validated using two representative model systems: a mixture of ammonium d-10 camphor sulfonate and polystyrene nanoparticles (PSNPs), and plasmonic gold helicoid nanoparticles, which exhibit both chiral absorption and scattering. For the former case, the CDS spectra show opposite signs to the CD spectra because the PSNPs are achiral scattering particles and the CDS spectra are affected by the chiral absorption. For the latter case, both CD and CDS spectra exhibit matched resonance wavelengths and stronger responses to the right-handed circularly polarized light, indicating that the chiral absorption and scattering arise from the same plasmonic resonance modes. To the best of our knowledge, this work represents the first experimental demonstration of the concurrent acquisition of ensemble-averaged CD and CDS spectra. The presented technique enables a direct and accurate comparison of CD and CDS spectra acquired under identical conditions.

\end{abstract}

\section{Introduction}
Chiroptical spectroscopy is a label-free and non-destructive technique for probing microscopic structural information through the interaction of circularly polarized light with matter. In particular, circular dichroism (CD), commonly defined as the differential absorption of left-handed circularly polarized (LCP) light and right-handed circularly polarized (RCP) light, has been widely employed to analyze the secondary structures of biomolecules and to differentiate enantiomers. Moreover, circular differential scattering (CDS) provides insights into structural and orientational information of biopolymers \cite{Maestre,Gregory}. It complements conventional CD spectroscopy for studying mesoscale chiral architectures and macromolecules \cite{Ni}.

CD typically arises from mechanisms such as exciton coupling, geometric phase retardation, or plasmonic resonances. Accurate CD measurements generally require sufficient optical transmission, which makes it challenging to implement in samples with low optical transparency \cite{Kelly, Pan}. In contrast, CDS exhibits exceptional sensitivity to long-range organized architectures extending up to the micrometer scale and is applicable to opaque or morphologically complex samples \cite{Savenkov}. Chiral complex systems, such as the enhancement of weak CD signals by plasmonic or dielectric nanostructures, biological macromolecules, and particulate biopharmaceutical samples, exhibit both chiroptical absorption and scattering \cite{Cai, Schäferling}. The combined observation of CD and CDS could offer deeper insights into the systems \cite{Zhang2019, Gratiet2020}.

 The RCP and LCP light is primarily implemented by rotating linear polarizers \cite{Tsai, Ding}, liquid crystal polarization modulators \cite{Zhou}, and photoelastic modulators (PEMs) \cite{ Diaspro, Shindo}. The first two approaches employ lower modulation frequencies. They are compatible with charge-coupled devices (CCDs) and typically rely on differential detection, with an ellipticity sensitivity limited to 30 mdeg level \cite{Reponen, Li}. The PEM approach is typically combined with lock-in detection, resulting in a simplified experimental setup and improved performance. By modifying a commercial CD spectropolarimeter equipped with a PEM, Gregory et al. characterized the CDS of chloroplasts and demonstrated its dependence on wavelength and scattering angle \cite{Gregory}. In recent years, several groups have reported single nanoparticle (NP) CDS measurements \cite{Gratiet2020, Wang}. However, the high numerical aperture objectives used in microscopy may introduce artifacts, such as polarization distortion, leakage of linear dichroism and the fixed orientation effect of the NPs \cite{Spaeth}. 

Although several single-function instruments are capable of measuring either CD or CDS, simultaneous acquisition on the same sample remains challenging. When the two signals are obtained from separate setups, differences in light sources and electronic components introduce systematic errors \cite{Quatela, Yan}. This limitation becomes more pronounced in dynamic processes such as protein aggregation and NP growth, where experimental conditions, including temperature and stirring speed, inevitably change between sequential measurements. Our dual-channel design addresses these issues, as both channels share the same light source, PEM, and electronic components. The parallel configuration further enables highly time-synchronized acquisition of CD and CDS from a single sample under identical conditions. In addition, conventional CD instruments usually measure the chiral extinction (circular differential light intensity attenuation due to both absorption and scattering) rather than true absorption. The presented dual-channel instrument may provide a powerful tool for decoupling chiral absorption from the chiral extinction.

In addition to detailing the instrument design and calibration, this work employs ellipticity to characterize both CD and CDS signals. The functionality of the instrument is subsequently validated through two distinct case studies: a mixture of polystyrene nanoparticles (PSNPs) and ammonium d-10 camphor sulfonate (a mixture of achiral scatterers and chiral absorbers), and chiral plasmonic gold NPs (exhibiting both chiral absorption and scattering). The former case represents a scenario related to chirality-enhanced biological systems \cite{Ma2023}. The latter are promising NPs exhibiting advantageous chiral properties \cite{Lee2018, Kuzyk}.

\begin{figure*}[!t]
\centering
\includegraphics[width=5in]{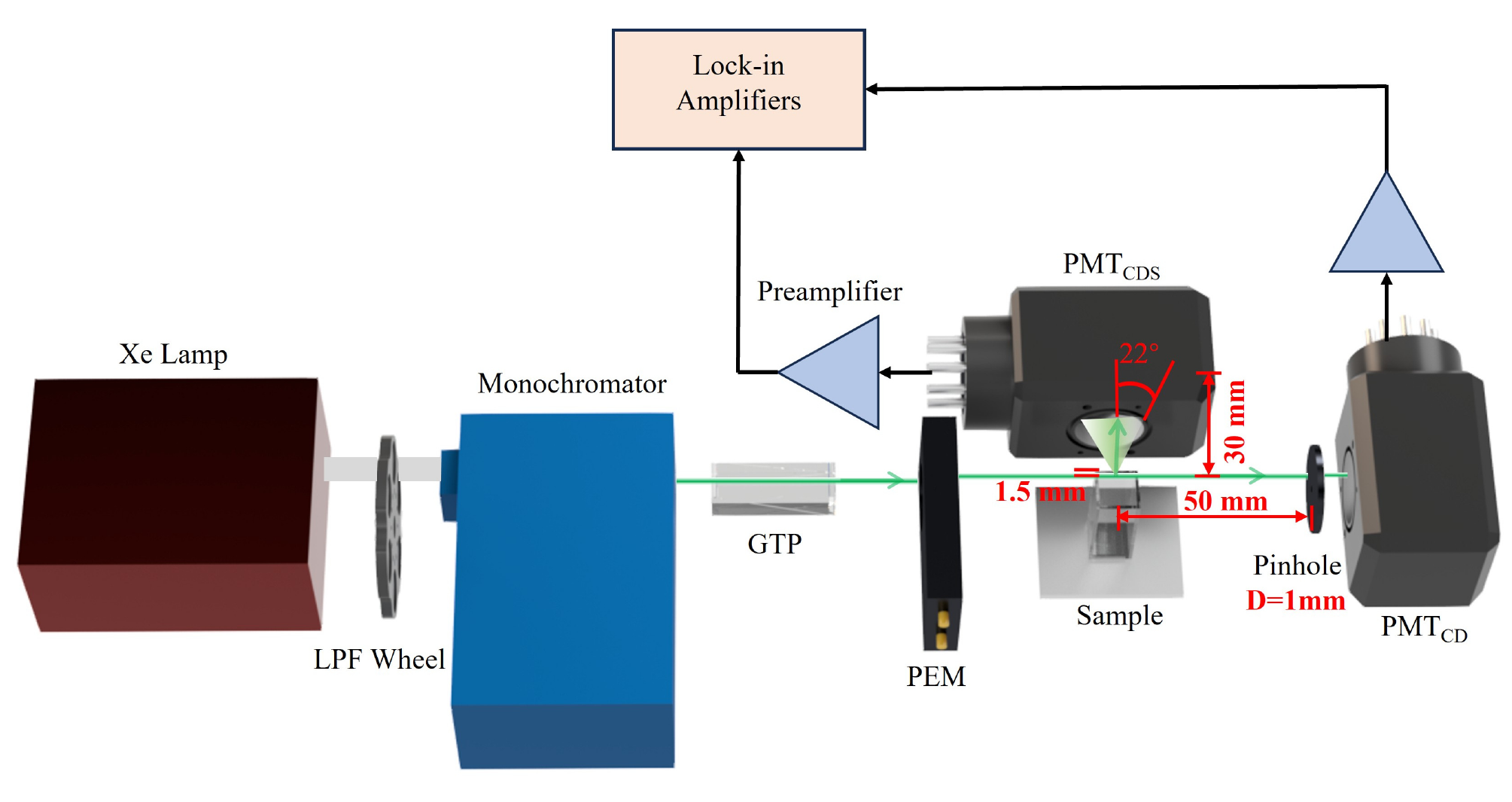}
\caption{{Schematic diagram of the dual-track chiral spectrometer, LPF: longpass filter, PEM: photoelastic modulator, PMT: photomultiplier tube, and GTP: Glan–Thompson polarizer.}}
\label{Fig1}
\end{figure*}

\section{Instrumentation design}

\subsection{System Integration}

A dual-channel chiral spectrometer for simultaneous CD and CDS measurements is implemented using a PEM, lock-in detection, and a grating-based monochromator, achieving an improved signal-to-noise ratio (SNR) through polarization modulation and coherent detection.  Additionally, prism-based double monochromators \cite{Rodger} or grating-based double monochromators\cite{ Shindo1998} are typically used to reduce the stray light within the system and therefore enhance the detection sensitivity. Here, we aim to demonstrate a more accessible and cost-effective configuration. Accordingly, a grating-based single monochromator is employed, and its performance limit will be explored.

As shown in Fig. 1, a 300 W Xe lamp (Sciencetech, LH-S-300X) serves as the broadband light source, and its output is directed through a 25 cm focal length monochromator (Sciencetech, 9055). The beam is re-collimated and passes through a mechanical shutter (Thorlabs, SH1). A Glan–Thompson polarizer (OptoSigma, GTPB-15-32.5SN) then generates a linearly polarized beam oriented at $45^\circ$ relative to the optical axis of the PEM, providing a polarization extinction ratio of less than $5\times10^{-6}$ and a large effective aperture of 15 mm. A PEM with sinusoidally varying quarter-wave retardation is used to modulate the light into alternating LCP and RCP at 42 kHz. The light then passes through a 10 mm path-length cuvette, which is placed in a temperature-controlled cuvette holder. A 1 mm diameter pinhole is positioned in front of the photomultiplier tube (PMT$_{\text{CD}}$, Hamamatsu R955P) to reduce the acceptance half-angle to less than $1^\circ$, thereby minimizing the scattering and stray light on the CD detector. The PMT$_{\text{CDS}}$ is oriented at $90^\circ$ with respect to the excitation light propagation direction. Owing to the large active area of PMT$_{\text{CDS}}$ (8 mm $\times$ 24 mm), the scattering horizontal collection half-angle is $22^\circ$. It should be noted that the excitation beam (1 mm in width) is positioned 1.5 mm from the exit side wall of the cuvette to minimize the influence of chiral absorption on the CDS measurement. The configuration of the two channels is the same to facilitate direct comparison of CD and CDS spectra. 

The PMT$_{\text{CDS}}$ can also be positioned above the sample holder, enabling the measurement of films or powders. Both CD and CDS signals are amplified by current preamplifiers. Each resultant signal is demodulated by a lock-in amplifier, yielding the corresponding AC and DC components.  The ellipticity is ultimately derived from the two components. 

\subsection{Ellipticity and dissymmetry factor}
The observable physical quantities from the photodetectors are the transmitted and scattered light intensities. The CD ellipticity is defined by
\begin{equation}
\label{Eq1}
\tan(\theta^{CD}) = \frac{E_R - E_L}{E_R + E_L}
\end{equation}
where $E_R$ and $E_L$ are the electric field amplitudes of the RCP and LCP light, respectively. When $\lvert E_R - E_L \rvert \ll (E_R + E_L)$, the ellipticity is
\begin{equation}
\label{Eq2}
\theta^{CD} \approx \frac{E_R - E_L}{E_R + E_L} = \frac{\sqrt{I_R} - \sqrt{I_L}}{\sqrt{I_R} + \sqrt{I_L}}
\end{equation}
where $I_R$ and $I_L$ represent the intensities of the RCP and LCP light. Assuming $I_R \approx I_L$, the ellipticity can be further simplified as

\begin{equation} 
  \label{Eq3}
\theta^{CD} \approx\frac{1}{2}\cdot\frac{I_R-I_L}{I_R+I_L}
\end{equation} 
Given that the measurement range in this work is $\left|\frac{I_R - I_L}{I_R + I_L}\right|\leq 0.2$, the resulting relative ellipticity error with the assumption $I_R \approx I_L$ is no more than 1\%.

The transmitted intensities are $I_R=I_0{10}^{-A_R}$ and $I_L=I_0{10}^{-A_L}$, respectively, where $A_R$ and $A_L$ are absorbance of the RCP and LCP light. In this work, “absorbance” is used in a generalized sense to describe total optical attenuation at the detector, including contributions from both true absorption and scattering. We obtain
    \begin{equation}
    \label{Eq4}
    \theta^{CD}(rad)=\frac{1}{2}\frac{I_0(10^{-A_R}-10^{-A_L})}{I_0(10^{-A_R}+10^{-A_L})}
   \end{equation}
 By defining $\Delta A=A_L-A_R$, we have  
   \begin{equation}
    \label{Eq5}
    \theta^{CD}(rad)=\frac{1}{2}\frac{1-10^{-\Delta A}}{1+10^{-\Delta A}}
   \end{equation}
   When $\Delta A$ is small, using the first-order Taylor series expansion, we have    
\begin{equation}
    \label{Eq6}
    \theta^{CD}(rad)\approx\frac{ln10}{4}\Delta A 
   \end{equation}
From the measurement perspective, the excitation light is modulated by PEM, the transmitted light is expressed as  
\begin{equation}
	\label{Eq7}
	\begin{split}
		I_{\text{det}} = \frac{1}{2} I_0 \big[ &(1 + \sin(\delta_0 \sin \omega t)) 10^{-A_R} \\
		&+ (1 - \sin(\delta_0 \sin \omega t)) 10^{-A_L} \big]
	\end{split}
\end{equation}
where $\delta_0$ is the peak retardance, $\omega$ is the modulation frequency.

As shown in Fig.1, $I_{\text{det}}$ is detected by the PMT and converted to a voltage signal by the preamplifier. The output of the preamplifier signal is given by 

\begin{equation}
	\label{Eq8}
	\begin{split}
		V_{\text{det}} = &\frac{1}{2} eQGI_0 (10^{-A_R} + 10^{-A_L}) \\
		&+ \frac{1}{2} eQGI_0 (10^{-A_R} - 10^{-A_L}) \sin(\delta_0 \sin \omega t)
	\end{split}
\end{equation}
where $e$ is the electron charge, $Q$ is the photocathode quantum efficiency and $G$ is the total gain by the PMT and preamplifier. Then the signal is fed into a lock-in amplifier. The first and the second terms in Eq. (8) are related to the DC and AC components of the output of the lock-in amplifier, respectively. We therefore obtain \cite{Drake} 
\begin{equation}
	\label{Eq9}
	\theta^{CD}(rad)=\frac{ln10}{4}k\frac{V_{AC}}{V_{DC}}
\end{equation}
$k$ is the scale factor and mainly depends on the PEM and lock-in configurations. Subsequently, we convert the $\theta^{CD}$ in the unit of mdeg  \cite{Kuroda}  

\begin{equation}
	\label{Eq10}
	\theta^{CD}\left(mdeg\right)=32982k\frac{V_{AC}}{V_{DC}}
\end{equation}
Since the differential absorbance is derived from transmitted light while the CDS signal is obtained directly from scattered light, the absorbance and scattering spectra have opposite sign conventions. The $\theta^{\text{CDS}}$ is therefore expressed as 
\begin{equation}
	\label{Eq13}
	\theta^{\text{CDS}} = \frac{1}{2} \frac{ I'_L-I'_R}{I'_L+I'_R}
\end{equation}

By setting the lock-in amplifier to the same configuration as that of the CD channel, the detected signal of the CDS channel is given by

\begin{equation}
	\label{Eq11}
	\begin{split}
		V'_{\text{det}} = &\frac{1}{2} eQG (I'_R+I'_L) \\
		&+ \frac{1}{2} eQG ( I'_R-I'_L ) \sin(\delta_0 \sin \omega t)
	\end{split}
\end{equation}
where variables with primes denote the quantities associated with CDS, maintaining physical definitions consistent with those used in the CD derivation. We have 

\begin{equation}
    \label{Eq13}
   \theta^{\text{CDS}} (\text{rad}) = -\frac{1}{2} k \frac{V'_{\text{AC}}}{V'_{\text{DC}}}
   \end{equation}

Subsequently, we convert the $\theta^{CDS}$ to the unit of mdeg

\begin{equation}
    \label{Eq14}
   \theta^{\text{CDS}} (\text{mdeg}) = -32982 \frac{2k}{\ln 10} \frac{V'_{\text{AC}}}{V'_{\text{DC}}}
   \end{equation}

The dissymmetry factor, or $g$ factor, is also widely used to characterize the chirality of samples. The $g$ factor of CD is commonly defined as $g^{\text{CD}} = \frac{\Delta A}{A}$. Meanwhile, two conventional definitions of $g^{\text{CDS}}$ have been reported, including $g^{\text{CDS}}=\frac{I_L' - I_R'}{I_L' + I_R'}$\cite{Maestre, Gratiet2020} and $g^{\text{CDS}}=\frac{2(I_L' - I_R')}{I^{'}_L + I_R'}$ \cite{Wang, Xie}. The $g^{\text{CD}}$ is expected to be independent of concentration according to the Beer–Lambert law. As this work aims to investigate the chiral response across various sample concentrations, the ellipticities $\theta^{CD}$ and $\theta^{CDS}$ are employed to characterize the samples. 
\begin{figure}[!t]
\centering
\includegraphics[width=3.5in]{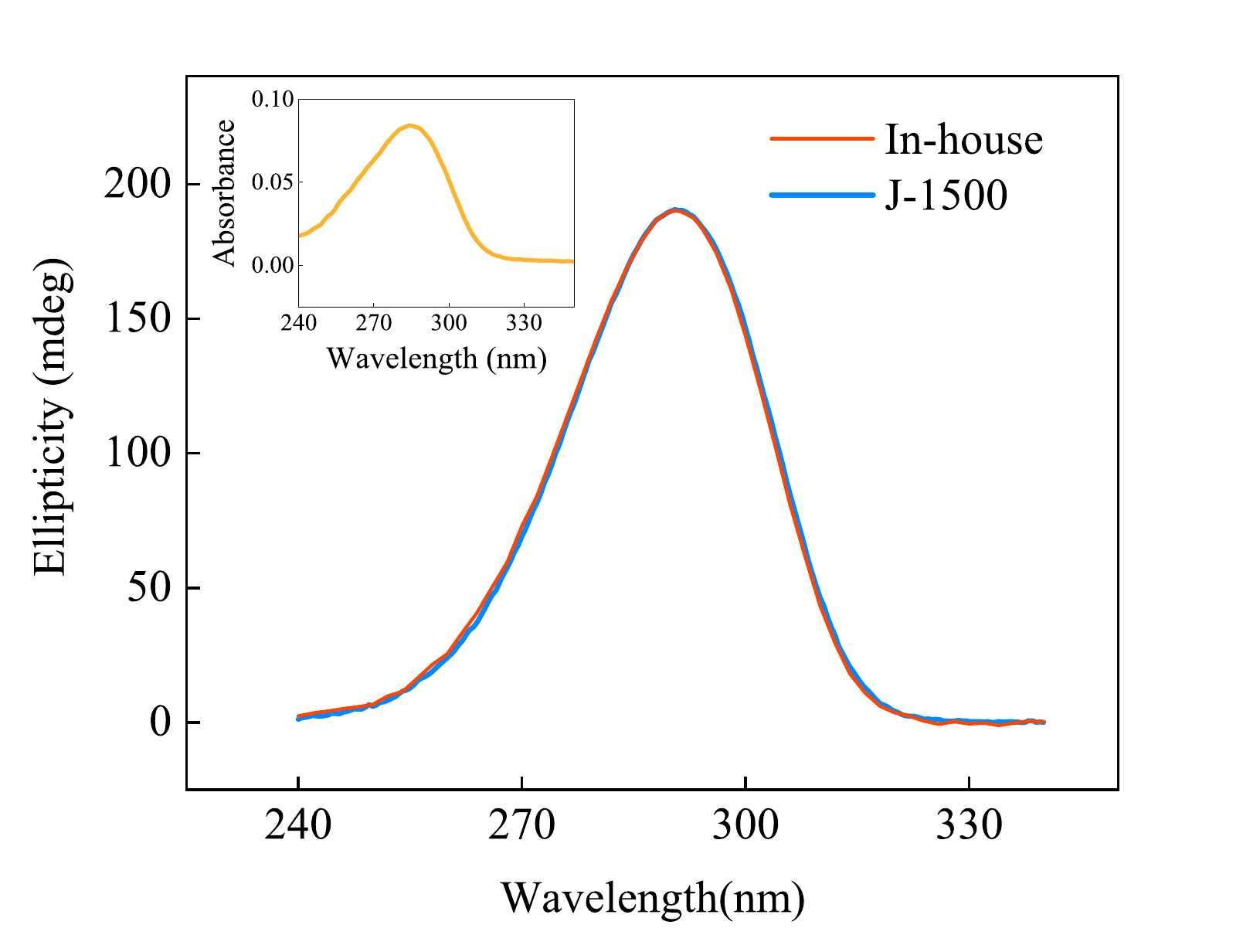}
\caption{Performance validation of the in-house CD system, which demonstrates high consistency with a commercial J-1500 spectrophotometer. The inset displays the absorbance spectrum of the ACS sample. Both CD spectra are measured with a digital integration time of 4 s.}
\label{Fig2}
\end{figure}
\begin{figure*}[!t]
\centering
\includegraphics[width=5in]{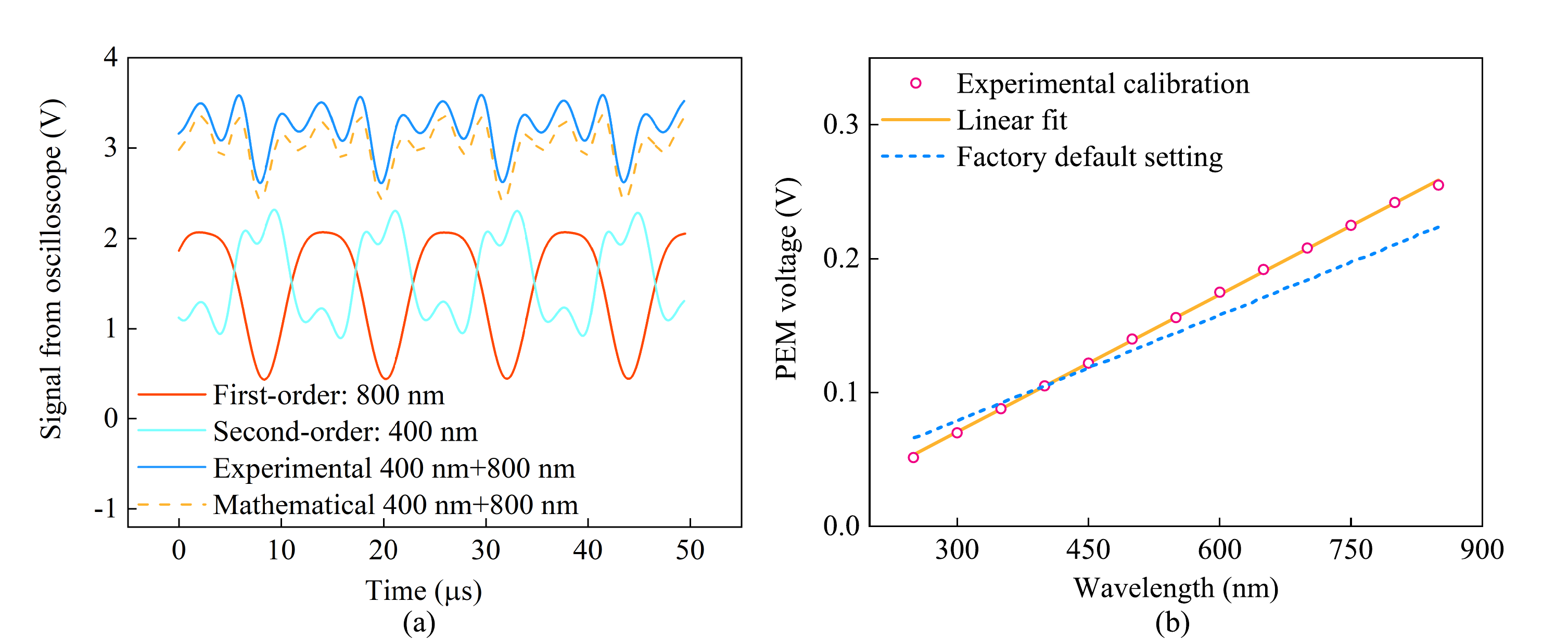}
\caption{(a) Influence of second-order diffraction on the calibration of PEM retardation. (b) Calibration of PEM driving voltage to maintain constant retardation.}
\label{Fig3}
\end{figure*}
\section {Calibration and characterization}
\subsection{Wavelength and scale calibration}
Careful calibration and characterization are essential for the developed analytical tool. Wavelength calibration is implemented by aligning the monochromator with an Hg/Ar lamp (Newport, 6035). The monochromator is equipped with two gratings with blaze wavelengths of 300 nm and 500 nm, respectively. Each grating has a groove density of 1200 lines/mm. The first grating is designated for operation in the range of 240–400 nm, and the second covers 400–850 nm. By setting the bandwidth to about 0.5 nm, we implement a multi-point calibration and achieve a wavelength accuracy of $\pm 0.2$ nm across the entire spectral range.

The scale calibration is performed with a certified ammonium d-10 camphor sulfonate (ACS) standard. The sample is dissolved in deionized water using a 10 mm path-length cuvette. The concentration is 0.06\% (w/v) and is expected to have an ellipticity of $190.4 \pm 1$ mdeg \cite{Rodger}. We first finely adjust the concentration by UV-Vis absorption spectrum, based on the standard molar absorptivity of 34.5 M$^{-1}$cm$^{-1}$ at the absorption maximum of 285 nm \cite{Chen1977}. The ellipticity is then measured using the CD channel. From these measurements, we determine that the correction factor $k$ equals 1.2. The CDS channel is calibrated to the same scale factor as the CD due to both channels sharing the same PEM and lock-in configurations.

The wavelength and scale calibration results are shown in Fig. 2. It also presents a comparison between our in-house CD system and a commercial CD spectrophotometer (Jasco J-1500). The peak wavelengths agree well at 291 nm, and the corresponding CD amplitudes differ by approximately 0.1\%. Meanwhile, ten repeated measurements are performed for both the CD and CDS channels using a mixture of ACS and PSNPs (chiral absorbers mixed with the achiral scatterers). The statistical peak amplitudes are 637.1 $\pm$ 0.5 mdeg for CD and -381.7 $\pm$ 0.9 mdeg for CDS, demonstrating high repeatability in both channels.

\subsection{PEM calibration}
The PEM utilizes a piezoelectric element to periodically squeeze and release an optical crystal, inducing a time-dependent birefringence that modulates the polarization of the incident light. The retardation calibration requires placing the PEM between a pair of crossed linear polarizers oriented at $+45^\circ$ and $-45^\circ$, and with the PMT serving as the photodetector. The signal is demodulated by a lock-in amplifier. Typically, the driving voltage scales linearly with the wavelength. The half-wave retardation is calibrated based on the oscilloscope calibration method, which identifies the appropriate driving voltage from the appearance of a flat-top waveform \cite{ Wang2019, Oakberg}.

The second-order diffraction from the grating distorts the waveform pattern. Taking the 800 nm wavelength as an example, we separate the first-order and the second-order components by using different kinds of optical filters. As shown in Fig. 3(a), the first-order diffraction passing through a 500 nm longpass filter produces a flat-top waveform. However, the second-order diffraction, passing through a filter with the center wavelength of 400 nm and a bandwidth of 10 nm, produces a significantly distorted waveform. When the excitation light directly goes through the PEM, it produces multiple overlapping peaks and makes it difficult to calibrate the PEM retardation. Therefore, an order-sorted filter wheel is positioned in front of the monochromator, which consists of three long-pass filters and automatically changes with the wavelength to avoid the second-order diffraction impact. Fig. 3(b) shows the in-house and factory default driving voltages, both of which exhibit good linearity. The difference in their slopes is attributed to variations in the beam shape and divergence \cite{Sutherland}.

\subsection{Baseline}
To understand the baseline, we first describe the process for maintaining a constant DC component. Due to sample absorption, the non-uniformity of the excitation light spectrum, and the wavelength-dependent quantum efficiency of the PMT, the photon flux incident on the detector varies significantly across a broadband spectrum. As the electronic system operates within a finite voltage range, a fuzzy control algorithm is implemented to prevent signal saturation. This algorithm automatically adjusts the PMT high voltage at each wavelength to stabilize the DC signal at 1 V. 

The baseline is a systematic offset unrelated to the sample. As shown in Fig. 4 (a)(b), we measure the CD baselines and the PMT voltages with different concentrations of  PSNPs. Despite the PMT voltage changes due to the scattering of PSNPs, the CD baselines are identical and on the order of mdeg. Therefore, the CD baseline could be efficiently removed by the conventional method, which uses the buffer solution \cite{Jones2024}.

The detected signal of the CDS channel is typically much weaker than that of the CD channel due to the 90$^{\circ}$ detection geometry. When the scattered photon flux is lower, the PMT will be driven at higher voltages to maintain a 1V DC signal. As shown in Fig.4 (c) and (d), the baseline is highly related to the scattering intensity, indicating that matching the scattering spectrum with achiral particles could be a solution to correct the CDS baseline. We apply the strategy to 123 nm chiral plasmonic gold helicoids (hereafter referred to as Helicoids), as shown in Fig. 5. While 60 nm Au NPs exhibit a scattering peak at 540 nm, their scattering is significantly reduced in the 600--750 nm region. To improve the scattering spectral matching, we prepare a mixture of two Au NPs with scattering peaks at 540 nm and 650 nm, respectively. The resulting scattering spectrum and PMT voltage profile demonstrate improved agreement with those of the Helicoids. Consequently, the baseline correction obtained with the mixed NPs is more accurate, as further discussed in the next section.

\begin{figure}[!t]
\centering
\includegraphics[width=3.5in]{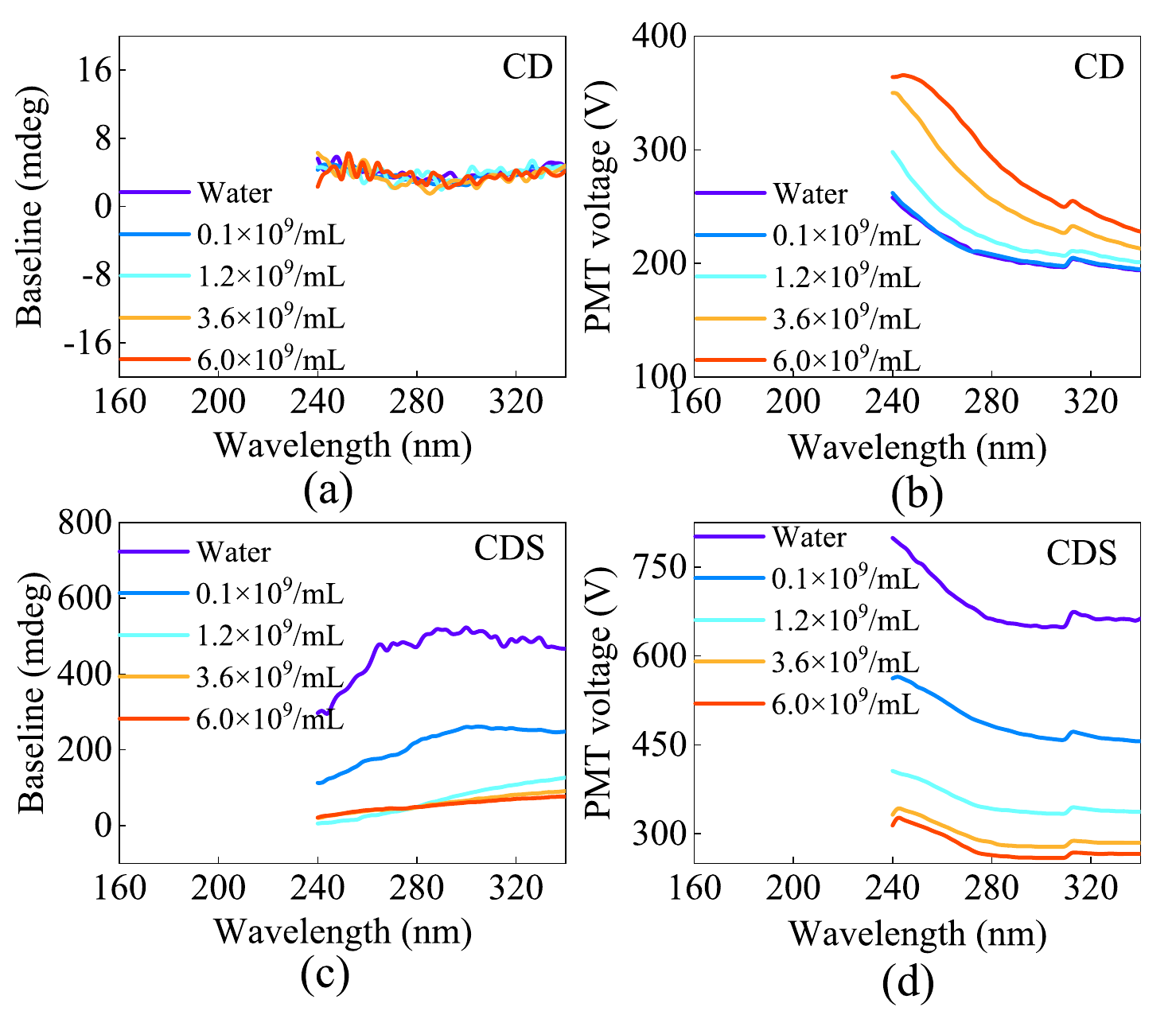}
\caption{Relationship between the baseline and PMT voltage, measured with PSNPs. (a),(b) Baseline and PMT voltage of CD, respectively. (c),(d) Baseline and PMT voltage of CDS. All data for each sample are acquired in a single measurement.}
\label{Fig4}
\end{figure}

\begin{figure*}[!t]
\centering
\includegraphics[width=5in]{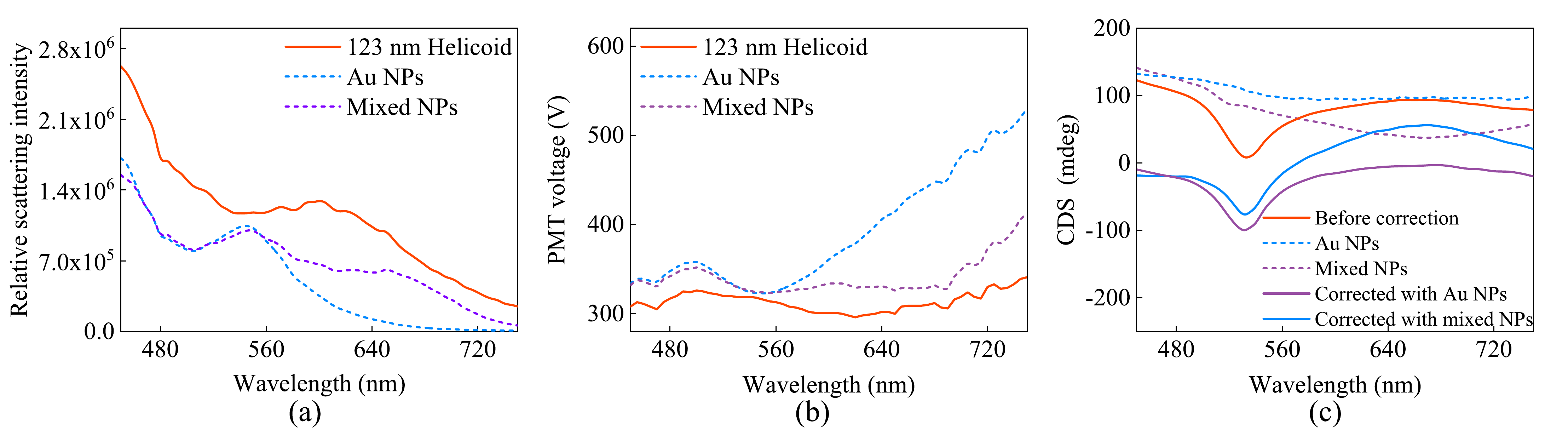}
\caption{(a) Scattering spectra measured by a fluorometer (Horiba, fluoroMax) in synchronous mode. (b) PMT voltages. (c) CDS baseline correction. Au NPs show a scattering peak at 540 nm, and mixed NPs are a combination of two types of Au NPs, with scattering peaks at 540 nm and 650 nm, respectively.}
\label{Fig5}
\end{figure*}

\subsection{{Noise floor}}
Noise floor affects the system's capacity to detect weak signals and is relevant to baseline fluctuations \cite{Castiglioni}. To improve the SNR, the system employs high-sensitivity PMTs exhibiting gains as large as $10^7$. The PEM is modulated at 42 kHz to suppress low-frequency interference, such as 60 Hz noise and its harmonics from utility power. Furthermore, the lock-in amplifier extracts the signal at a reference frequency while filtering out noise at other frequencies.  As shown in Fig.6, the amplitudes of the CD and CDS are 1.4 mdeg and -14 mdeg, respectively. The RMS noises for CD and CDS are 0.013 mdeg and 0.85 mdeg, respectively. These results are obtained using a mixture of 0.0005\% (w/v) ACS and $1.2\times 10^9$ NPs/mL PSNPs for the CD channel, and 0.0045\% (w/v) ACS mixed with $1.2\times 10^9$ NPs/mL PSNPs for the CDS channel.

Table I lists the noise budget of CD and CDS measurements. It is important to note that the bandwidth and sensitivity of the lock-in amplifier are set to 0.8 Hz and 10 mV, respectively. Meanwhile, we use a 40 s integration time during the data acquisition, so the white frequency noises, such as shot noise and Johnson noise, and the quantization noise due to the digital bit depth are effectively suppressed. As a result, the fluctuation of PEM residual birefringence is believed to be the primary contributor to the CD noise floor, which arises from the temperature dependence of the residual birefringence\cite{wang2009}. It can be further mitigated by integrating the instrument in a sealed enclosure \cite{Qiang2025}.

\begin{figure}[!t]
\centering
\includegraphics[width=3.5in]{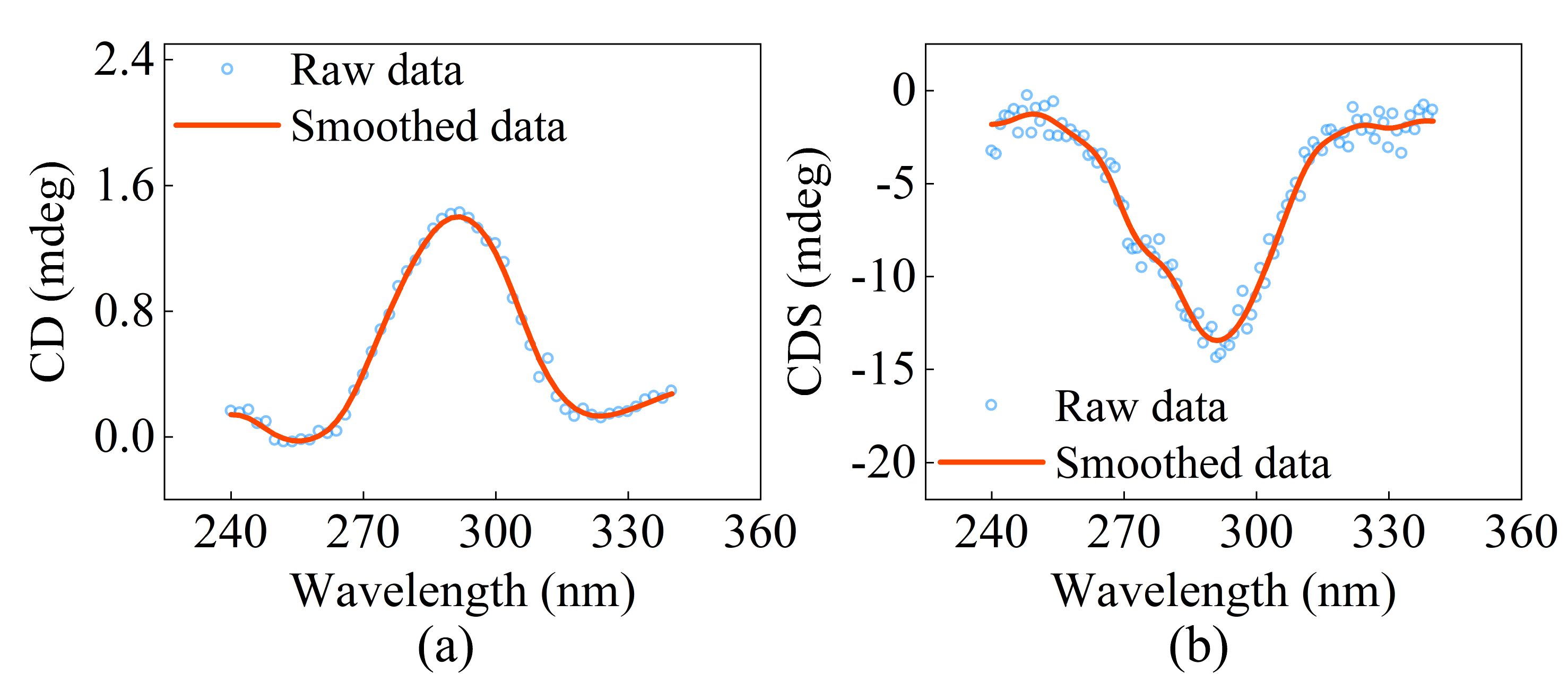}
\caption{ (a) CD and (b) CDS spectra measured separately with ACS mixed with PSNPs samples. The PSNPs concentration of the samples is constant at $ 1.2\times 10^9$ NPs/mL,  while the ACS concentrations for CD and CDS measurements are set to be approximately 0.0005\% (w/v) and 0.0045\% (w/v), respectively. The spectra are measured with an integration time of 40 s and smoothed by the digital FFT filter.}
\label{Fig6}
\end{figure}
The CDS baseline at 291 nm reaches 58 mdeg, which is more than four times the signal amplitude of 14 mdeg. Due to the wide acceptance half-angle of the PMT$_{\text{CDS}}$ and the weak detected signal intensity (0.5 nW), the impact of the stray light becomes the dominant concern \cite{DiN2012}. It can be substantially suppressed by introducing a relay imaging system with a spatial filter.

\begin{table}[!t]

\renewcommand{\arraystretch}{1.3}
\caption{CD and CDS Noise Budget}
\label{tab:noise_budget}
\centering
\begin{tabular}{>{\raggedright\arraybackslash}p{2.8cm}>{\centering\arraybackslash}p{2.0cm}>{\centering\arraybackslash}p{2.0cm}}
\hline\hline
Noise type & CD noise (mdeg) & CDS noise (mdeg) \\
\hline
PEM residual birefringence       & 0.008                & 0.008  \\
Shot noise                       & 0.006                & 0.007  \\
PMT dark noise and nonlinearity  & $<1\times10^{-4}$    & 0.001  \\
Lock-in phase drift              & 0.002                & 0.002  \\
Preamplifier additive noise       & 0.004                & 0.004  \\
Digital data acquisition         & 0.003                & 0.003  \\
Stray light                      & 0.001                & 0.800  \\
\hline
Estimated total                  & 0.011                & 0.800  \\
Measured total                   & 0.013                & 0.850  \\
\hline\hline
\end{tabular}
\end{table}

To illustrate the current capabilities and limitations of our instrument, Table II summarizes key characteristics alongside those of a commercial CD instrument. Meanwhile, to the best of our knowledge, no commercial CDS instrument is currently available. Therefore, we include the transient full-Stokes measurements for the reference. This technique demonstrates state-of-the-art performance with high time resolution and sensitivity, and it could be extended to ensemble-averaged CDS measurements \cite{Reponen}. The wavelength range of the current implementation is 240--850 nm. It can be extended to the deep-UV by upgrading the light source and implementing N$_2$ purge. The spectral coverage can also be extended to the near-infrared region by replacing multialkali PMTs with InGaAs photodetectors. Notably, this chiroptical spectrometer employs a single-grating monochromator, demonstrating that high-performance chiroptical measurement can be achieved with a cost-effective optical configuration.

\begin{table}[!t]

\renewcommand{\arraystretch}{1.3}
\caption{Comparison of Specifications Between the Presented Instrument and Existing CD/CDS Systems}
\label{tab:comparison}
\centering
\begin{tabular}{>{\raggedright\arraybackslash}p{1.9cm}>{\centering\arraybackslash}p{2.8cm}>{\centering\arraybackslash}p{2.1cm}>{\raggedleft\arraybackslash}p{2.8cm}}
\hline\hline
 & Jasco 1500 & Transient full-Stokes method~\cite{Reponen} & This work \\
\hline
Wavelength range & 163--950 nm (standard) & 400--900\,nm & 240--850\,nm \\
Nitrogen purge & Yes & No & No \\
Wavelength accuracy & $\pm$0.5\,nm (500--800\,nm) & $\pm$0.2\,nm (500--800\,nm) & $\pm$0.2\,nm (500--800\,nm) \\
Modulation frequency & 50\,kHz & N/A & 42\,kHz \\
{RMS noise} & {CD: 0.007\,mdeg} & {$\sim$30\,mdeg} & CD: 0.013\,mdeg \\
 &  &  & CDS: 0.85\,mdeg \\
Scanning speed & 10000\,nm/min & Simultaneous, 2 ns time resolution& 20000\,nm/min \\
Detector & PMT & ICCD & PMT \\
\hline\hline
\end{tabular}
\end{table}
     
\section {Applications}
\subsection {Mixture of ACS and PSNPs}
A chiral sample (ACS) and an achiral scattering sample (PSNPs) are mixed as an aqueous suspension to validate the functionality of the developed instrument. The ACS concentration across all samples is held constant at 0.2\% (w/v). The size of the PSNPs is 200 nm and the concentrations of PSNP$_1$ and PSNP$_2$ are $1.2\times10^9$ NPs/mL and $3.6\times10^9$ NPs/mL, respectively.
 \begin{figure}[!t]
\centering
\includegraphics[width=3.5in]{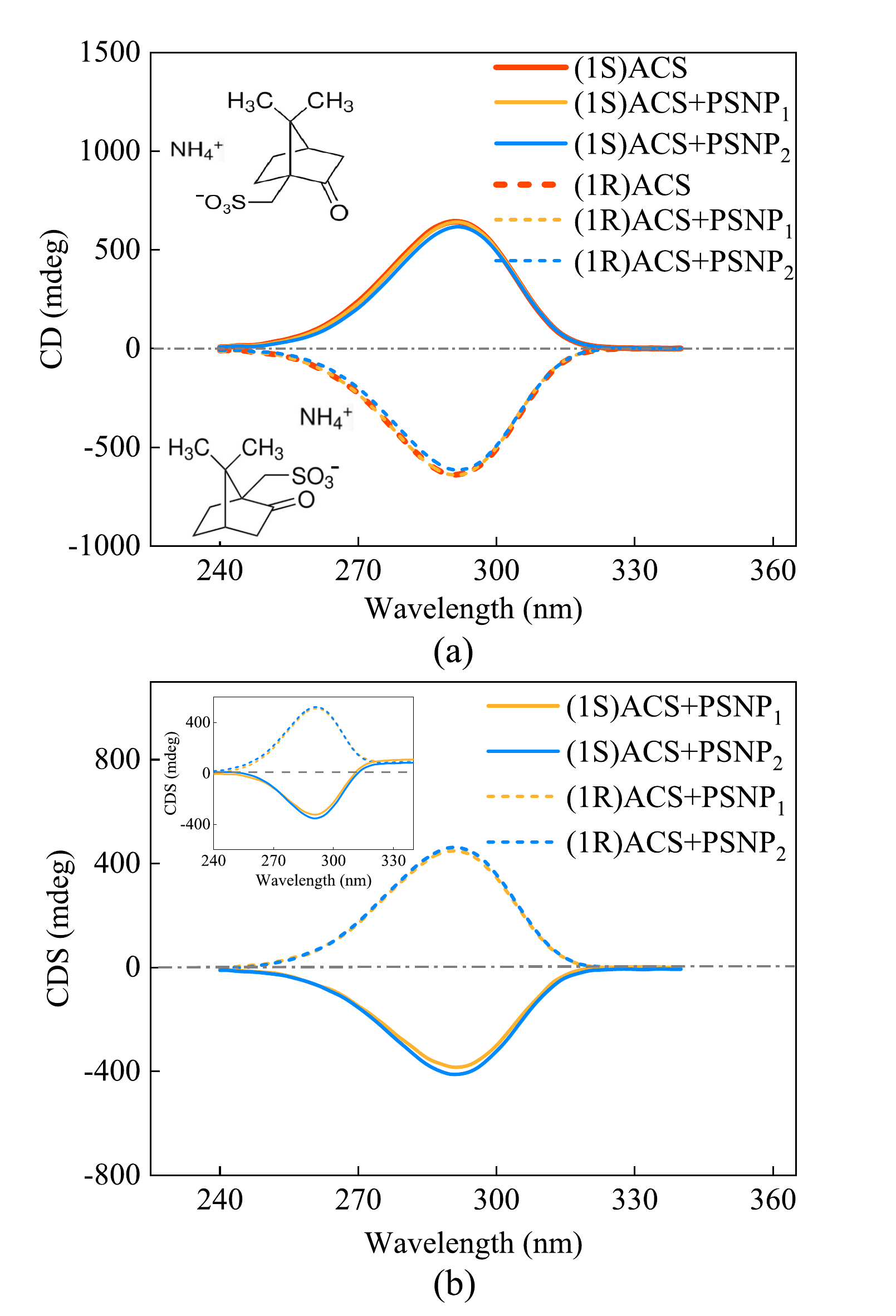}
\caption{(a) CD and (b) CDS of ACS and PSNPs mixture samples, which are baseline-corrected using PSNPs profiles of corresponding concentrations shown in Fig. 4. The ACS concentration across all samples is held constant at 0.2\% (w/v). PSNP$_1$ and PSNP$_2$ are the same material but different concentrations, which are $1.2\times10^9$ NPs/mL and $3.6\times10^9$ NPs/mL, respectively. The inset of (b) shows the raw data without baseline removal. Given that ACS samples are well-known pure absorbers and their CDS signal is unreal due to the extremely low scattering intensity (PMT voltage of around 650 V), CDS spectra of ACS-only are therefore not presented in Fig. 7(b).}
\label{Fig7}
\end{figure}

As shown in Fig. 7, the two enantiomers (1S)-ammonium camphor sulfonate (denoted as (1S)ACS) and (1R)-ammonium camphor sulfonate (denoted as (1R)ACS) exhibit good symmetry in their CD spectra. When adding PSNPs into the samples, the amplitudes and peak wavelengths remain consistent with those of the pure ACS samples. The CDS spectra exhibit the opposite signs to the CD signals. Given a positive CD signal, the absorbance of LCP light is greater than that of RCP light. Therefore, the scattering light intensity of LCP will be smaller than that of RCP, yielding a negative CDS signal. Moreover, the raw data exhibited slight distortions, which are baseline-corrected using PSNPs profiles of corresponding concentrations shown in Fig. 4(c). The final results for the enantiomers show highly symmetric profiles, indicating the effectiveness of the baseline correction.

Notably, the spectra detected by PMT$_{\text{CDS}}$ might be suspected to arise from the chiral absorption of the scattered light. This concern can be addressed by comparing the peak amplitudes of the CD and CDS spectra. The effective absorption path lengths for spectra detected by PMT$_{\text{CD}}$ and PMT$_{\text{CDS}}$ are 10 mm and 1.5 mm, respectively. If the absorbance dominated the PMT$_{\text{CDS}}$ detected spectra, the peak magnitude would be expected to be significantly lower than that of the CD spectra \cite{Wathudura}. Given that the observed CDS peak intensity is approximately two-thirds that of the CD signal, we conclude that the spectra detected by PMT$_{\text{CDS}}$ primarily arise from CDS, but with a minor impact from chiral absorption. In addition, we show that CD and CDS spectra exhibit the same sign at their peaks in the application of plasmonic helicoids. Two applications with distinct scattering properties validate the effective detection of the CDS.

\subsection {Chiral plasmonic gold helicoids}
Helicoids are synthesized using gold NPs as templates. Guided by glutathione, gold NPs preferentially grow along specific directions and form helically twisted structures. For Helicoids, surface plasmons are optically excited, and the light can be coupled into standing or propagating surface plasmon modes through helically twisted structures, which significantly enhances light-matter interactions by localized surface plasmon resonance (LSPR). Benefiting from the twisted structures and the resonant enhancement, Helicoids exhibit a strong chiroptical response \cite{Lee2018, Wan}. Depending on the handedness of glutathione employed during synthesis, they can be fabricated as either right-handed (D Helicoids) or left-handed (L Helicoids) enantiomers. Here, we investigate the D Helicoid with the sizes of 180 nm and 123 nm, which are dispersed in a 1 mM cetyltrimethylammonium bromide (CTAB) solution. Both samples are measured in a 10 mm path-length cuvette at a concentration of $2 \times 10^8$ NPs/mL \cite{Han2025}.

The chiroptical response of the Helicoids mainly originates from the interference between electric dipole and magnetic dipole moments \cite{Lee2018}. With statistical analysis of the SEM images, the gap width and depth of the 180 nm Helicoids are determined to be 25 nm and 65 nm, respectively. For the 123 nm Helicoids, the corresponding values are 20 nm and 45 nm. Numerical simulations further confirm that the 180 nm Helicoids, with more pronounced gaps, exhibit more distinct current loops and an enhanced magnetic dipole contribution \cite{Han2025}. Therefore, more pronounced chirality has been observed with 180 nm Helicoids. Furthermore, as the particle size increases, the phase of the incident electromagnetic field varies across the particles. This retardation effect leads to a red shift of plasmonic resonance \cite{Mayer2011}.

As shown in Fig. 8, the CD and CDS show consistent chiral response profiles and their peak wavelengths are 535 nm and 625 nm for the 123 nm and 180 nm Helicoids, respectively. Moreover, the CD and CDS spectra exhibit the same sign. It indicates that the Helicoids exhibit both stronger absorption and scattering on RCP light according to the definitions in Eqs. (3) and (11). Jointly considering the resonance wavelength, resonance bandwidth, and the signs of CD and CDS, it can be inferred that both chiral absorption and scattering originate from the same set of chiral plasmonic modes. These plasmonic modes are primarily governed by the interference between electric dipole and magnetic dipole moments, in agreement with the previous numerical simulations \cite{Xie,Zhang2025}.

 \begin{figure}[!t]
\centering
\includegraphics[width=3.5in]{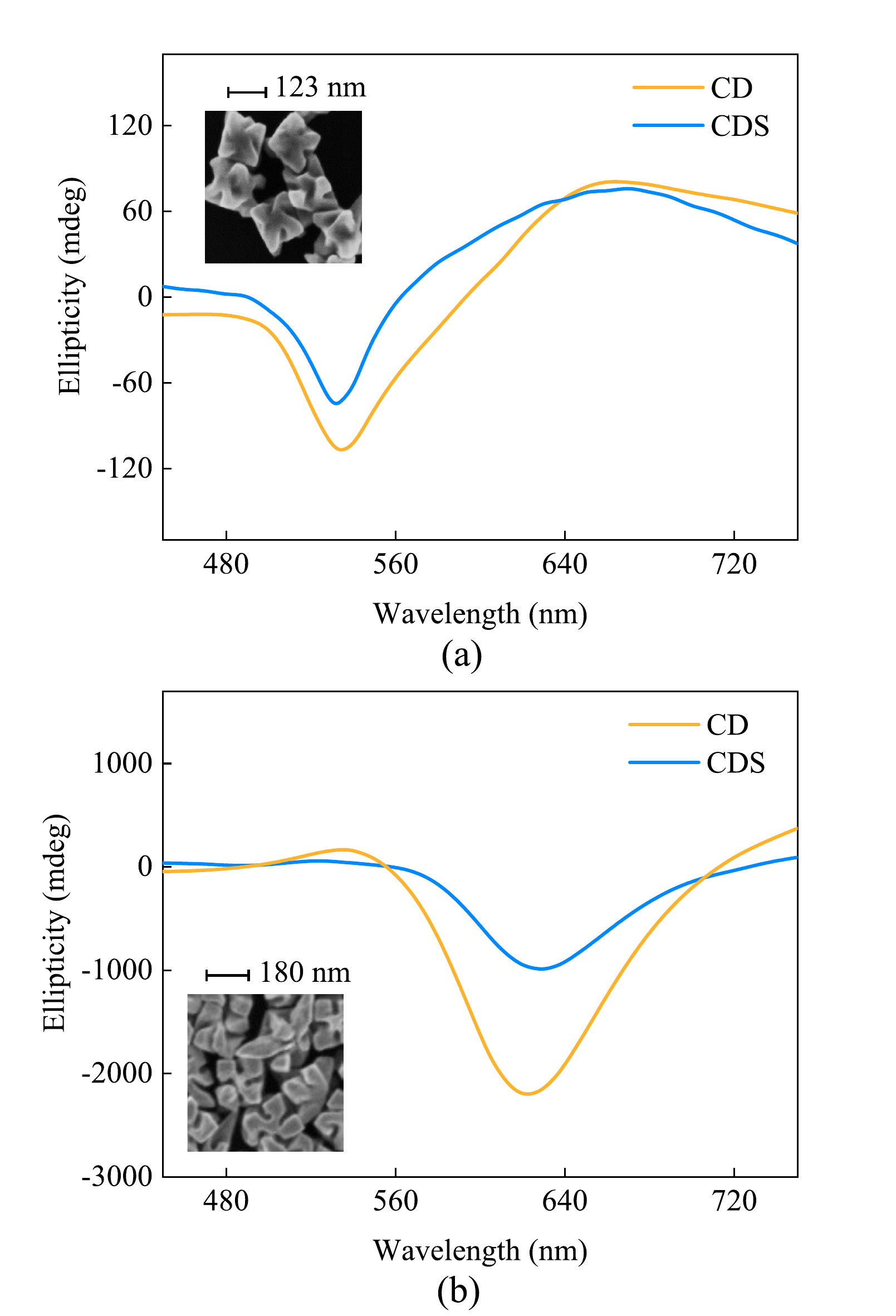}
\caption{CD and CDS spectra of (a) 123 nm and (b) 180 nm chiral plasmonic NPs. Insets display scanning electron microscope (SEM) images of the NPs.}
\label{Fig8}
\end{figure}

\section {Conclusions and outlook}
In summary, we developed a dual-channel chiroptical spectrometer, incorporating a PEM, a grating-based monochromator, and lock-in detection. To enable direct comparison of the CD and CDS spectra, we characterize both with ellipticity. The RMS noises of the CD and CDS are $0.013 \text{ mdeg}$ and $0.85\text{ mdeg}$, respectively. CD and CDS spectra of the mixture of ACS and PSNPs at different concentrations and enantiomers have been measured. The results exhibit symmetric and highly reproducible profiles, demonstrating the effectiveness and reliability of the measurements and baseline correction. The instrumental performance is also validated with chiral plasmonic gold Helicoids. Both CD and CDS spectra display matched resonance wavelength and spectral shapes, revealing that the resonance modes producing the CD and CDS are identical \cite{Han2025, Mark2013}. 

Since both CD and CDS signals arise from the interference between multiple electromagnetic multipoles, changes in the surrounding refractive index can modify not only the resonance wavelength but also the relative strength and spectral shape of the CD and CDS signals. In particular, shifts in the plasmon resonance and changes in the electromagnetic field distribution around the particle can alter the interference between electric dipole and magnetic dipole moments, which is relevant to the observed chiroptical effects \cite{Mayer2011}. Our instrument provides a new tool on this topic. Future research will explore the angular dependence of CDS spectra, which may provide deeper insights into the interpretation of CD and CDS spectra \cite{AO2021, Miles}.

\end{document}